\documentclass[a4paper,11pt]{article}
\pdfoutput=1 

\usepackage{jheppub} 

\usepackage[T1]{fontenc} 

\title{Higgs Inflation: constraining the top quark mass and breaking the $H_0$-$\sigma_8$ correlation}

\author[a]{Jamerson G. Rodrigues,}
\author[b,c]{Micol Benetti,}
\author[a]{Rayff de Souza}
\author[a]{and Jailson Alcaniz}

\affiliation[a]{Observatório Nacional, 20921-400, Rio de Janeiro, RJ, Brazil}
\affiliation[b]{Scuola Superiore Meridionale, Largo San Marcellino 10, 80138, Napoli, Italy}
\affiliation[c]{Istituto Nazionale di Fisica Nucleare (INFN) Sezione di Napoli, Complesso Universitario di Monte Sant'Angelo, Edificio G, Via Cinthia, I-80126, Napoli, Italy}

\emailAdd{jamersoncg@gmail.com}
\emailAdd{micol.benetti@unina.it}
\emailAdd{souzarayff@gmail.com}
\emailAdd{alcaniz@on.br}

\abstract{Extending previous results [JHEP 11 (2021) 091], we explore aspects of the reheating mechanism for non-minimal Higgs inflation in the strong coupling regime. We constrain the radiative corrections for the inflaton’s potential by considering the Coleman-Weinberg approximation and use the Renormalization Group Equations for the Higgs field to derive an upper limit on the quark top mass, $m_t$. Using the current Cosmic Microwave Background, Barion Acoustic Oscillation, and Supernova data, we obtain $m_t \leq 170.44$ GeV, confirming the observational compatibility of the model with recent $m_t$ 
 estimates reported by the CMS collaboration. We also analyze the breakdown of the well-known correlation involving the Hubble constant $H_0$ and the clustering parameter $\sigma_8$, which makes the model interesting in light of the cosmological tensions discussed over the last decade.}


\begin{document} 
\maketitle
\flushbottom

\section{Introduction}
\label{Sec:1}

The fundamental theory behind the initial conditions that led to the temperature fluctuations in the Cosmic Microwave Background (CMB) \cite{Aghanim:2019ame,Ade:2018gkx} and the formation of Large-Scale Structure (LSS) of the universe \cite{Beutler:2011hx,Alam:2016hwk,Scolnic:2017caz} remains an open question in modern cosmology. In this context, the paradigm of inflation rises as the most elegant description of the primordial Universe \cite{Guth:1980zm,Linde:1981mu,Albrecht:1982wi,Mukhanov:1981xt,Starobinsky:1982ee}. In order to induce cosmic acceleration, the dynamical equations for the inflaton field must enable a slowly varying solution, leading to a quasi-de Sitter Universe. In the well-known slow-roll mechanism this is achieved in an approximately flat direction of the inflaton's scalar potential. 

One particularly appealing approach is to induce a non-minimal coupling between the inflaton and gravity, which results in a plateau at the large field regime \cite{Starobinsky:1980te,Spokoiny:1984bd,Salopek:1988qh} and drives the model predictions to the sweet-spot of CMB observations \cite{Aghanim:2018eyx}. From the phenomenological perspective, one specially interesting model was introduced by Berzrukov and Shaposhnikov \cite{Bezrukov:2007ep}, where the standard Higgs field rules the inflationary period at early times. Such configuration allows one to compare the predictions of the model for the cosmological observables with the phenomenology of the related particles at electroweak scale of energy. Such analysis was explored in a number of interesting papers, see e.g.  \cite{Barvinsky:2008ia,GarciaBellido:2008ab,Bezrukov:2014bra,Lee:2018esk,Rodrigues:2021txa}.  

Although robust, the analysis of inflationary models rely on a set of assumptions about the evolution of cosmological quantities. In particular, the evolution of cosmological scales from the moment they cross the Hubble radius during inflation up to the their re-entrance at later times must be matched to all the eras of the cosmological expansion in order to solve the horizon problem \cite{Liddle:2003as}. The matching condition can be written in the form
\begin{equation}
    \ln{\left[\frac{k}{a_0 H_0}\right]} = -N_k-N_{rh} - N_{RD} + \ln{\left[\frac{a_{eq}H_{eq}}{a_0 H_0}\right]} + \ln{\left[\frac{H_{k}}{H_{eq}}\right]},\label{Eq:01}
\end{equation}
where $N_k$ is the number of e-folds the universe expanded between the horizon crossing moment of the pivot scale $k$ and the end of inflation and $N_{rh}$ is the number of e-folds counted from the end of inflation to the onset of the radiation dominance in the early Universe (reheating). Also, $N_{RD}$ gives the amount of expansion between the end of reheating and the end of radiation dominated era, while the subscript ``$eq$" and ``$0$" represent quantities evaluated at matter-radiation equality and the present, respectively. One is not able to set the amount of expansion the universe experienced in the inflationary period, $N_k$, without further information about the subsequent periods of the expansion. This is particularly problematic for the reheating period.

In a previous communication \cite{Rodrigues:2021txa}, we performed a Monte-Carlo Markov Chain (MCMC) analysis of CMB and clustering data to check the observational viability of non-minimally coupled $\phi^4$ models for a fixed inflationary e-fold number. In particular, we considered the first order correction to the perturbative expansion of the inflationary potential, also known as Coleman-Weinberg approximation \cite{Coleman:1973jx}, and constrained possible radiative corrections coming from the underlying field theory supporting this cosmological scenario. In addition, we used the two-loop Renormalization Group Equations to connect the model's predictions at inflationary energy scales to the electroweak observables and derived an estimate of the top quark mass $m_t$, indicating a possible tension  with the Monte-Carlo Tevatron and LHC reconstruction \cite{Zyla:2020zbs}. 

In this work, we extend and complement the analysis reported in \cite{Rodrigues:2021txa} by exploring the predictions of non-minimal Higgs inflation for a wide range of the inflationary e-fold number $N_{k}$ and, consequently, of $N_{rh}$. Following the procedure developed in \cite{Rodrigues:2021txa,Rodrigues:2020dod}, we employ a MCMC analysis  to compare the predictions of this inflationary scenario with the most recent Cosmic Microwave Background (CMB), Baryon Acoustic Oscillation (BAO), and Supernova (SN) data \cite{Aghanim:2019ame,Ade:2018gkx,Beutler:2011hx,Alam:2016hwk,Scolnic:2017caz}. In particular, we obtain new constraints on the radiative corrections coming from the underlying field theory supporting this cosmological scenario and derive an upper limit for the top quark mass, which is compared with recent $m_t$ measurements from different experiments. Furthermore, we also explore whether this model could shed some light on the so-called cosmological tensions, which include the well-known $H_0$ tension, a
$\sim 4\sigma$-discrepancy between direct measurements of $H_0$
using low-$z$ SN ($H_0 = 73.48 \pm 1.66 $ km/s/Mpc \cite{Riess:2018uxu}) and the $H_0$ estimate from current CMB data assuming the standard model ($H_0 = 67.72 \pm 0.41 $ km/s/Mpc \cite{Aghanim:2018eyx}) \cite{DiValentino:2021izs, DiValentino:2020zio}. It is worth mentioning that most of the
usual mechanisms to solve this problem have failed so far, as alleviating the $H_0$ discrepancy worsens the agreement of other parameters with the data. In particular, the clustering parameter, $\sigma_8$, is constrained at  $\sigma_8 = 0.766^{+0.024}_{-0.021}$ by the  Kilo-Degree Survey (KiDS-1000) lensing estimation \cite{KiDS:2020suj} and its correlation with the Hubble constant leads to significantly too high values as the value of $H_0$ increases. Breaking such a correlation is not only tricky but also 
challenging for many cosmological scenarios.

This work is organized as follows. In Sec.~\ref{sec:2}, we briefly introduce the non-minimal inflationary scenario and present the results of the slow-roll analysis. In Sec.~\ref{sec:3}, we discuss aspects of the reheating stage following the Higgs inflation and present the main results of our statistical analysis of the cosmological data. Sec.~\ref{sec:4} discusses the constraints derived on the top quark mass and some  implications on the current cosmological tensions. The main conclusions of this work are presented in Sec.~\ref{conclusions}.

\section{Non-minimal Inflation and Slow-Roll Analysis}
\label{sec:2}

As mentioned earlier, a common method to achieve slow-roll inflation is to induce a non-minimal coupling between the inflaton field and gravity. Such procedure yields non-canonical terms for the original scalar field and the metric, suggesting the use of a set of conformal transformations in order to obtain the theory description in the familiar Einstein-Hilbert formalism. A more detailed exposition of this approach can be found in \cite{Rodrigues:2021txa}. 

The Einstein frame lagrangian reads 
\begin{equation}
    \mathcal{L}_E = -\frac{M_P^2 \Tilde{R}}{2} + \frac{1}{2}(\partial_\mu \chi)^\dagger(\partial^\mu \chi) - V_E(\chi)\;,
\end{equation}
and the subsequent time evolution is dictated by the inflaton's potential 
\begin{equation}
    V_E(\chi) = \frac{\lambda M_P^4}{4\xi^2}\left(1 - e^{-\sqrt{\frac{2}{3}}\frac{\chi}{M_P}}\right)^2\left(1 + a'\ln\sqrt{\frac{1}{\xi}e^{\sqrt{\frac{2}{3}}\frac{\chi}{M_P}} - \frac{1}{\xi}}\right)\label{potential}
\end{equation}
where the large field regime is assumed for the inflaton, $\chi \gg \sqrt{6}M_P$, and a large coupling regime is assumed for the non-minimal coupling, $\xi \gg 1$. Note that the deviation from the tree level potential is quantified by the parameter $a' \equiv \beta_\lambda/\lambda$, where $\beta_\lambda$ is the running equation for the quartic coupling $\lambda$. The above potential was obtained by adopting the prescription II procedure to compute the radiative corrections in the Jordan frame and all couplings are computed at the scale $M = M_P$, where $M_P$ is the reduced Planck mass \cite{Rodrigues:2021txa}.

Once with the effective potential in the Einstein frame, the relevant slow-roll inflationary parameters can be readily computed, which can be related to the spectral index and tensor-to-scalar ratio, characteristic of the power spectrum of CMB perturbations probed by Planck \cite{Aghanim:2019ame}. Although the field strength $\chi_*$, necessary to compute the relevant inflationary parameters, cannot be measured directly, we can infer its value from the duration of inflation from horizon crossing up to the end of inflationary expansion, characterized by the number of e-folds, which is also dependent on the form of the potential (\ref{potential}).

\begin{figure}[t]
    \centering
    \includegraphics[scale = 0.78]{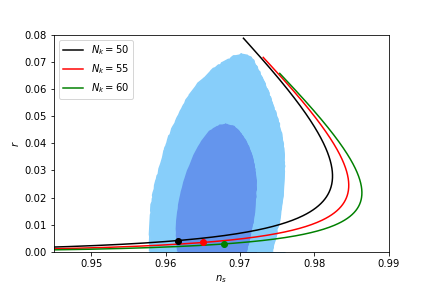}
    \caption{$n_s$ vs. $r$ for $N_k = 50, 55$ $\&$ $60$. The points in each curve indicate the parameters for a null resultant of the radiative corrections ($a' = 0$). The blue areas show the favored regions by \textit{Planck 2018}, with 68\% and 95\% confidence level (Planck TT, TE, EE + lowE + lensing + BK15 + BAO data set) \cite{Aghanim:2018eyx}.}
    \label{fig:1}
\end{figure}

However, the inflationary number of e-folds is not a free parameter entirely, as it is tied to the subsequent evolution of the universe, given its association with the horizon exit of relevant cosmological scales. Therefore, the relevant scales probed by Planck seem to correspond to an interval of 50-60 e-folds \cite{Liddle:2003as}, which guides our range of exploration of the parameter $N_k$.

In Fig. \ref{fig:1} we present our results for the spectral index and tensor-to-scalar ratio in the $n_S \, \times \, r$ plane,  with $a^\prime$ ranging from $-0.1$ (lower limit) to $1.0$ (upper limit)\footnote{The values of $a^\prime$ varying between [-0.010, 0.053], [-0.020, 0.036] and [-0.027, 0.023], corresponding to $N_k = 50$, $55$ and $60$, respectively, are in agreement with the 95\% C.L. Planck result.}. Note that there is a significant dependence of the inflationary predictions with the amount of expansion during inflation, achieving compatibility with the Planck result\footnote{This agreement relies on the slow-roll approximations for the inflationary parameters and the phenomenological power-law expansion of the primordial power spectrum.}. It is also important to mention that the results obtained for the prediction of inflationary parameters are highly independent of the coupling parameter $\xi$. 

\section{Reheating analysis and results}
\label{sec:3}

Between the end of inflation and the onset of a radiation-dominated universe, the universe undergoes a reheating period. Even though there are a number of proposals for the dynamics of the cosmos in this period \cite{Abbott:1982hn,Dolgov:1982th,Albrecht:1982mp,Kofman:1994rk,Traschen:1990sw,Kofman:1997yn,Shtanov:1994ce,Amin:2014eta}, the reheating era is exceptionally difficult to be constrained by observations, given the small length scales characteristic of this micro-physical process. For previous works exploring the impact of reheating to the cosmological observables see e.g. \cite{Dai:2014jja,Cook:2015vqa,Drewes:2015coa,Mishra:2021wkm} and references therein.

In order to understand the influence of the reheating period on the inflationary predictions, one can follow the steps developed in \cite{Cook:2015vqa} and resume the matching condition (\ref{Eq:01}) to the expression:
\begin{equation}
    N_{k} = \frac{-1+3\omega_{rh}}{4} N_{rh} - \ln{\left( \frac{V^{1/4}_{end}}{H_k} \right)} + 61.55\;, 
\end{equation}
where the amount of expansion through the inflationary period is explicitly related to the reheating characteristics of the proposed model. Here, $\omega_{rh}$ represents the effective equation-of-state parameter of the cosmological fluid during reheating, $V_{end}$ is the amplitude of the inflaton's potential energy at the end of inflation, $H_k$ is the Hubble parameter evaluated at horizon crossing and $k = 0.05$ Mpc$^{-1}$ is the pivot scale. We also consider $g_{rh} \sim 100$  for the relativistic degrees of freedom to obtain the numerical factor above. 

\begin{table}
    \centering
\begin{tabular}{|l|c|c|c|c|c}
\hline
 & $a'$ & $r_{0.02}$ & $H_0$ & $\sigma_8$ \\
\hline
$N_k$=50    & $0.179 \pm 0.072$ &  $0.032 \pm 0.013$ & $68.82 \pm 0.38$ & $0.841 \pm 0.005$ \\
$N_k$=52  & $0.040 \pm 0.015$ &  $0.007 \pm 0.002$ & $68.31 \pm 0.41$ & $0.835 \pm 0.005$ \\
$N_k$=54  & $0.011 \pm 0.014$ &  $0.004 \pm 0.001$ & $67.71 \pm 0.45$ & $0.817 \pm 0.003$ \\
$N_k$=54.5  & $0.009 \pm 0.013$ &  $0.004 \pm 0.001$ & $67.68 \pm 0.43$ & $0.811 \pm 0.003$ \\
$N_k$=55  & $0.010 \pm 0.013$ &  $0.004 \pm 0.001$ & $67.71 \pm 0.44$ & $0.804 \pm 0.003$ \\
$N_k$=56     & $0.022 \pm 0.015$ &  $0.005 \pm 0.001$ & $67.94 \pm 0.45$ & $0.793 \pm 0.003$ \\
$N_k$=58     & $0.283 \pm 0.169$ &  $0.044 \pm 0.019$ & $68.37 \pm 0.39$ & $0.779 \pm 0.004$ \\
$N_k$=60     & $0.243 \pm 0.088$ &  $0.042 \pm 0.015$ & $68.46 \pm 0.38$ & $0.766 \pm 0.005$ \\
\hline
\end{tabular}
\caption{Constraints for fixed $N_k$ at $68\%$ C.L. using the  Planck $TT,TE,EE + lowE +lensing +  BICEP2/Keck + BAO +Pantheon$ combination.}
    \label{tab:results}
\end{table}

In what concerns non-minimal inflationary models, it is possible to show that the inflaton condensate starts the reheating process oscillating with an effective matter-like equation of state ($\omega_1 = 0$) and, after crossing a critical value $\chi_{cr}$, finishes the process as a radiation-like component of energy ($\omega_2 = 1/3$) \cite{Bezrukov:2008ut,Gong:2015qha}. 
After some algebraic manipulations and using the approximation $H_k \sim \sqrt{V_*/3}$, valid during inflation, one obtains:
\begin{equation}
    N_k = -\frac{1}{4}N_1 - \ln{\left( \frac{V^{1/4}_{end}(a')}{\sqrt{V_*(a')/3}} \right)} + 61.55\label{eq:N1}
\end{equation}
where we highlight the $a'$ dependence of the inflationary potential.

We analyze the present model for fixed values of $N_k$ and compute the values of $V_{end}$ and $H_{k}$ following the slow-roll approximations. In our analysis we assume a standard cosmological model with a modified primordial spectrum in which the radiative correction parameter, $a'$, is free to vary. For the parameter estimation we use the free available  CosmoMC code~\cite{Lewis:2002ah}\footnote{This is a MCMC code interfaced with the Boltzmann solver Code for Anisotropies in the Microwave Background (CAMB)~\cite{Lewis:1999bs}. We modified CAMB following the indications of ModeCode~\cite{Mortonson:2010er, Easther:2011yq} in order to analyse the specific form of the potential $V(\phi)$.} and a combination of early and late data\footnote{We use the CMB Planck (2018) likelihood \cite{Aghanim:2019ame}, using Plik temperature power spectrum, TT, and HFI polarization EE likelihood at $\ell \leq 29$; BICEP2 and Keck Array experiments B-mode polarization data \cite{Ade:2018gkx}; BAO measurements from 6dFGS ~\cite{Beutler:2011hx}, SDSS-MGS~\cite{Ross:2014qpa}, and BOSS DR12~\cite{Alam:2016hwk} surveys, and the Pantheon sample of Type Ia supernovae \cite{Scolnic:2017caz}.} (for more details we refer the reader to \cite{ Rodrigues:2021txa}).
Table~\ref{tab:results} shows the derived constraints on the most significant parameters of our analysis. 

\begin{figure}[t]
    \centering
    \includegraphics[scale = 0.75]{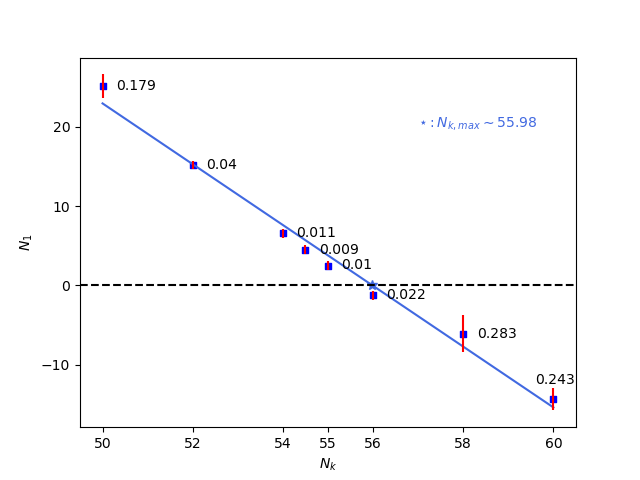}
    \caption{$N_k$ vs. $N_1$ for each inflationary number of e-folds taken into consideration. $N_1$ is given by the matching equation (\ref{eq:N1}), with $a'$ coming from the MCMC analysis (highlighted beside each point). Through a linear regression between the points (solid blue line),
we estimate a maximum number $N_k$ - where the transition to a radiation-dominated Universe happens instantaneously.}
    \label{fig:reh}
\end{figure}

Note that by computing the values of $V_{end}$ and $H_{k}$, we can obtain the corresponding values for $N_1$, i.e, the amount of expansion that the universe went through, as matter-like dominated, during the reheating process. The corresponding values are presented in Figure \ref{fig:reh}.
Note also that, for an expansion of $\sim$ 56 e-folds or greater during inflation, $N_1$ would have to assume negative values to satisfy the matching equation (\ref{eq:N1}). By definition, this condition would imply in a contraction of the universe between the end of inflation and the onset of the radiation-dominated epoch\footnote{It is also possible to obtain $N_1 > 0$ even for $N_k > 56$ if one considers exotic scenarios for the transition to radiation dominance, including intermediary phase transitions of the reheating fluid to an exotic component of energy $\omega^\prime >1/3$.}. Thus, following the standard approach, we discard these possibilities as non-physical. Therefore, we can tighten the bounds on the maximum value for the inflationary number of e-folds, which yields an instantaneous transition to the radiation-dominated expansion.

The results presented above are insensitive to the specific physical process that leads to the transition between matter and radiation-like expansion in the reheating. As pointed out in \cite{Bezrukov:2008ut,GarciaBellido:2008ab}, non-perturbative processes may occur before the perturbative decays become viable (preheating), displacing the transition between the two expansion behaviors, which is particularly true in the model of Higgs Inflation. In this context, a specially interesting result was obtained in \cite{Sfakianakis:2018lzf}, where the authors discussed the resonant production of Higgs and gauge degrees of freedom in the linear regime of the Higgs Inflation scenario. For $100<\xi<1000$, the preheating dominant process is the Higgs self-resonance, leading to $N_1 \simeq 3$. For higher values of the non-minimal coupling, $\xi>1000$, it was pointed out that a substantial amount of energy stored in the inflaton condensate is transferred to relativistic gauge bosons already at the very first oscillation of the background (instant preheating), leading to $N_1 = 0$. Note that these results are in agreement with our analysis for $N_k \simeq 55$ and $N_k \simeq 56$, respectively, which is also in agreeement with the MCMC result for the radiative corrections in the interval $a^\prime \simeq \left[-0.003, 0.037\right]$ at $68\%$ (C.L.).

\section{Physical and cosmological consequences} 
\label{sec:4}

\subsection{Constraints on the top quark mass}

It is helpful to recall that the result mentioned above is obtained in the framework of the Higgs Inflation scenario, where $a^\prime$ is associated with the $\beta$-function of the Higgs quartic coupling $\lambda$. Once the renormalization group equations for the standard Higgs couplings are considered, it is possible to link the cosmological constraints to the phenomenology of the associated particles at the electroweak scale of energy\footnote{The parameters considered in the definition of $a^\prime$ are evaluated at the renormalization scale $M=M_P$.}. In this context, following the approach developed in \cite{Rodrigues:2021txa}, one shall infer an upper limit on the top quark pole mass, $m_t \leq 170.44$ GeV, to reproduce the values of $a^\prime$ above. Also, it is worth emphasizing that this limit on $m_t$ is relatively insensitive to the amplitude of the non-minimal coupling once the strong limit ($\xi \gg 1$) is assumed.

The most precise constraints on the top quark mass are extracted from the kinematic reconstruction of the $t\bar{t}$ events where $m_t$ is employed in the Monte-Carlo generator in order to fit the data \cite{CDF:1994vkk,D0:2004rvt}. This MC top quark
mass is usually assumed to be the pole mass even though the theoretical
uncertainties inherent to this association are hard to quantify \cite{Bezrukov:2014ina}. From \cite{Workman:2022ynf}, the average value for the top quark mass is set to $m_t = 172.69 \pm 0.30$ GeV, obtained from LHC and Tevatron data. If contrasted with the limit on $m_t$ obtained from the cosmological analysis, this represents a significant discrepancy of $7.5\sigma$.

Instead, one may consider theoretically cleaner the inference of the top quark pole mass from the measurements of the cross-section of the top quark production, since the theoretical computation of $\sigma(t\bar{t})$ is explicitly performed in a renormalization scheme (e.g., $\overline{\text{MS}}$) \cite{Langenfeld:2009wd}. In this case, the average value obtained from the Tevatron and LHC runs is $172.5 \pm 0.7$ GeV \cite{Workman:2022ynf}, lowering the discrepancy with our cosmological estimate of $m_t$ to $\approx 3\sigma$. More recently, the CMS collaboration reported $m_t = 170.5 \pm 0.8$ GeV, obtained from the differential cross-section of the top production \cite{CMS:2019esx}. Such result perfectly agrees with the results of our cosmological analysis of the Higgs Inflation.

 \begin{figure}[t]
\centering
\includegraphics[width=0.7\textwidth]{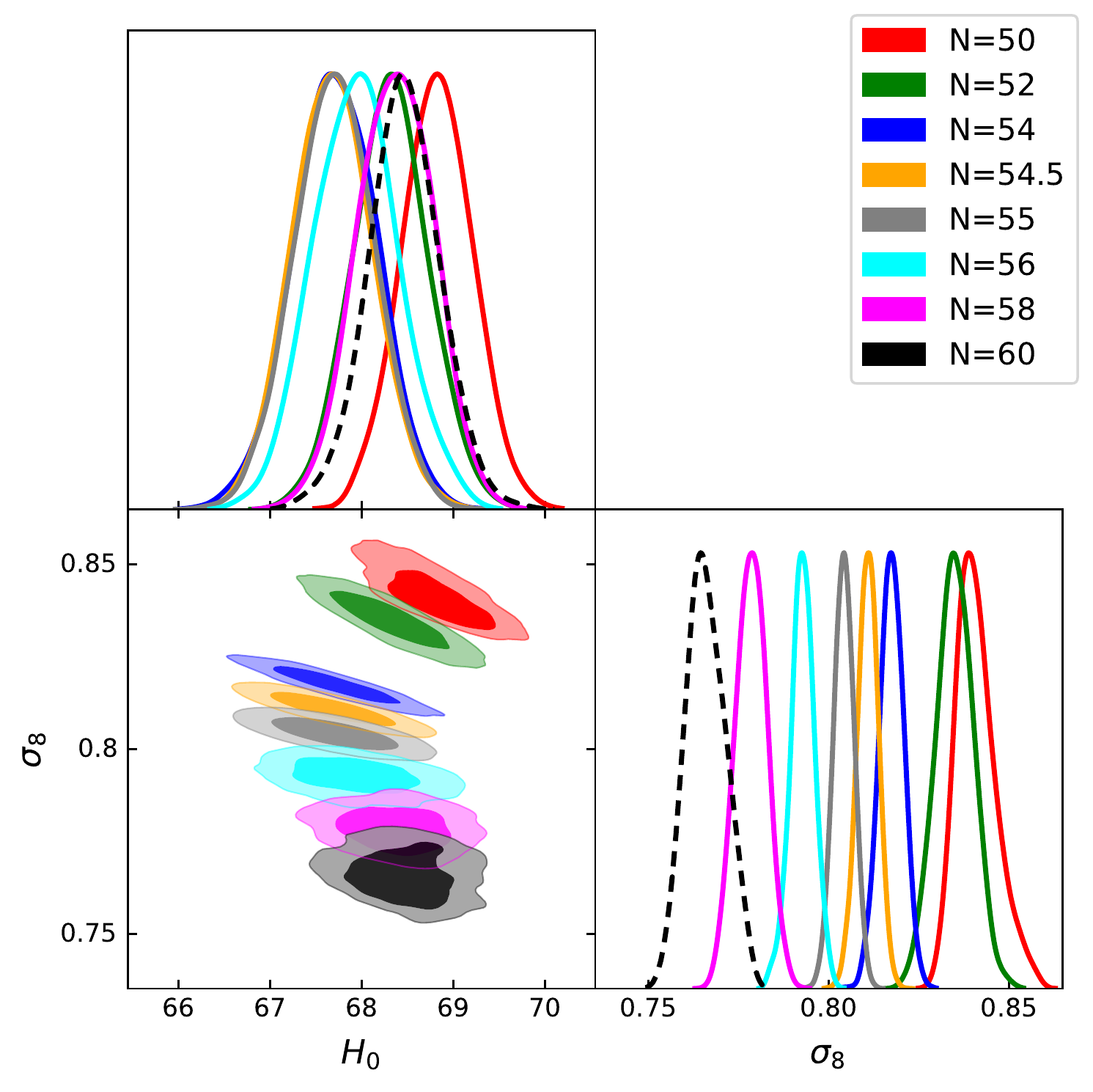}  
\caption{Confidence levels and posterior distributions for the $H_0$ and $\sigma_8$ parameters using the joint data set CMB Planck (2018) + BICEP2 and Keck Array + BAO + Pantheon SNe Ia sample and considering several values of $N_k$.}
\label{fig:H0-sigma8}
\end{figure}

 \subsection{The $H_0 - \sigma_8$ correlation}

The accuracy of cosmological and astrophysical measurements has significantly improved in recent decades. While this has led to increasingly evident confirmation of the validity of the standard cosmological model, it has also exposed some critical issues that have given rise to heated debate. The well-known $H_0$ tension 
has been extensively explored without concluding so far (we refer the reader to \cite{DiValentino:2021izs, DiValentino:2020zio} and references therein).

It has also been widely pointed out that some of the current attempts to solve the $H_0$ tension have failed because as they alleviate the discrepancy on $H_0$, they worsen the agreement of other parameters with the data. In particular, the clustering parameter, $\sigma_8$, is constrained at  $\sigma_8 = 0.766^{+0.024}_{-0.021}$ by the  Kilo-Degree Survey (KiDS-1000) lensing estimation \cite{KiDS:2020suj} and its correlation with the Hubble constant leads to values that are significantly too high as the value of $H_0$ increases. 

It is generally agreed that a model that manages to resolve both tensions is a model that breaks this degeneracy, but building such a model is proving difficult. So far, only a handful of scenarios seem to succeed, such as 
the conjecture of a universe transition from anti-de Sitter vacua to de Sitter vacua \cite{Akarsu:2022typ,PhysRevD.104.123512,PhysRevD.101.063528}, some early interacting models \cite{Reeves:2022aoi}
or  specific parametrizations of dark energy equation of state \cite{Naidoo:2022rda}. The model studied here is promising, as it breaks the degeneracy between the two parameters. In particular, Table \ref{tab:results} shows that as $N_k$ increases between the values of $50$ and $54.5$, the values of the radiative parameter, $a'$, $H_0$, and $\sigma_8$ decrease. Nevertheless, at the turning value of $N_k=54.5$, there is a behavior change, i.e., as $N_k$ increases, the values of $a'$ and $H_0$ also increase. In contrast, the value of the clustering parameter, $\sigma_8$, does not seem to be affected by this turning point and continues to decrease. It means that, for values of $N_k \in [54.5,60]$\footnote{As discussed earlier, we consider the cases $N_k > 56$ to be non-physical since they predict negative values of $N_1$.}, the correlation between $H_0$ and $\sigma_8$ breaks down, as also shown in Figure \ref{fig:H0-sigma8}. In particular, for the limiting value $N_k = 56$, i.e., an instantaneous transition to the radiation-dominated expansion, 
the degeneracy $H_0-\sigma_8$ is such that it reduces the $H_0$ tension, constraining $H_0= 67.94 \pm 0.45$ Km/s/Mpc, which is $\approx 3\sigma$ off from the SNe Ia measurements \cite{Riess:2018uxu} and allowing a value of $\sigma_8 = 0.793 \pm 0.003$, that is in full agreement with KiDS-1000 results~\cite{KiDS:2020suj}.

\section{Conclusions}
\label{conclusions}

In this work, we revisited the non-minimal inflationary scenario subject to radiative corrections. By performing an observational analysis of the $\phi^4$ primordial potential, non-minimally coupled to the Ricci scalar, in light of the most recent CMB, clustering and Supernova data and considering the allowed range for the observable inflationary e-folds, we constrained the possible values of the radiative corrections of the inflaton potential, encoded in the parameter $a^\prime$, and the usual set of cosmological parameters. 

From this analysis, we presented two main results. First, we set an upper limit to the number of e-folds from the horizon crossing moment up to the end of inflation, $N_k \lesssim 56$, relative to instantaneous reheating, by considering the matching equation for the pivot scale $k=0.05$ Mpc$^{-1}$. An even more stringent limit is imposed once considered the preheating structure of the Higgs Inflation, yielding $55 \lesssim N_k \lesssim 56$. Accordingly, the MCMC analysis of the model translates into an upper bound for the top quark pole mass, $m_t \leq 170.44$ GeV, which raises two possible interpretations for the consistency of the model at low-energies. For example, considering the value of the top quark mass reconstructed from the analysis of LHC and Tevatron data, $M_t = 172.69 \pm 0.30$ GeV \cite{Zyla:2020zbs}, implies a significant tension of $7.5\sigma$ between the observed low-energy value and the amount inferred by the cosmological MCMC analysis. On the other hand, assuming the top quark mass extracted from differential cross-section of the top production, $M_t = 170.5 \pm 0.8$ GeV, obtained by the CMS collaboration \cite{CMS:2019esx}, we found a perfect agreement between the cosmological analysis of the Higgs field and its electroweak behaviour. 

Second, the MCMC analysis of current observational data confirms the observational viability of the model and shows that for the interval $N_k \in [54.5,60]$, it can break down the well-known $H_0 - \sigma_8$ correlation (see Table 1). In particular, considering an instantaneous transition to the radiation-dominated expansion, which occurs for $N_k = 56$, the $H_0$ tension is reduced to $\approx 3\sigma$ whereas the value of $\sigma_8$ shows a complete agreement with KiDS-1000 results. 

These results reinforce the need to investigate Higgs inflation and its extensions from both theoretical and observational sides and show that perspectives for a complete coherence of the scenario may converge once data from future collider experiments \cite{AbdulKhalek:2018rok,Azzi:2021gwg} improve our understanding of the physics at the eletroweak scale. 

\acknowledgments
We thank Andr\'e Sznajder for helpful conversations. JGR acknowledges financial support from the Programa de Capacita\c{c}\~ao Institucional (PCI) do Observat\'orio Nacional/MCTI. MB acknowledges Istituto Nazionale di Fisica Nucleare (INFN), sezione di Napoli, iniziativa specifica QGSKY. RdS is supported by the Coordena\c{c}\~ao de Aperfei\c{c}oamento de Pessoal de N\'ivel Superior (CAPES). JSA is supported by CNPq (Grants no. 310790/2014-0 and 400471/2014-0) and Funda\c{c}\~ao de Amparo \`a Pesquisa do Estado do Rio de Janeiro FAPERJ (grant no. 233906). We also acknowledge the use of CosmoMC and ModeCode packages. This work was developed thanks to the use of the National Observatory Data Center (CPDON).


\bibliographystyle{JHEP}
\bibliography{cas-refs}



\end{document}